\documentclass[12pt]{iopart}

\usepackage{graphicx} 
\usepackage{epstopdf}                
\usepackage{bm}
\usepackage{iopams}
\usepackage{setstack}

\newcommand{\PRA}{{\it Phys. Rev.} A }
\newcommand{\PRB}{{\it Phys. Rev.} B }

\newcommand{\OC}{{\it Opt. Commun.} }
\newcommand{\JOSAB}{{\it J. Opt. Soc. Am.} B }

\begin{document}

\title[Bose-Hubbard Mode Splitter]{Entanglement via a three-well Bose-Hubbard system and via an optical beamsplitter}

\author{ M.~K.~Olsen}

\address{The University of Queensland, School of Mathematics and Physics,
Brisbane, Queensland 4072, Australia}

\date{\today }

\begin{abstract}

We compare and contrast the entangling properties of a three-well Bose-Hubbard model and an optical beamsplitter. The coupling between the different modes is linear in both cases, and we may identify two output modes.  Obvious differences are that our Bose-Hubbard model, with only the middle well initially occupied, does not have a vacuum input port, there is no equivalent of a collisional, $\chi^{(3)}$ nonlinearity with the beamsplitter, and the results of the Bose-Hubbard model show a time-dependence. In the non-interacting case, we obtain analytic solutions and show that, like a beamsplitter, the Bose-Hubbard system will not produce entanglement for classical initial states. We also show that whether inseparability or entanglement are detected depends sensitively on the criteria measured, with different criteria giving contradictory predictions.

\end{abstract}

\pacs{03.75.Gg,03.65.Ud,67.85.Hj}

\submitto{\JPB}

\ead{uqmolsen@uq.edu.au}

\maketitle

\section{Introduction}
\label{sec:intro}

In this article we extend previous work which combined 
the two fields of quantum information and ultra-cold bosons to propose a method for the fabrication of spatially isolated entangled atomic populations~\cite{CVCandme}. We do this by comparing the performance of a three-well Bose-Hubbard system~\cite{BHmodel1,BHmodel2} to that of an optical beamsplitter for the production of spatially separated output modes, using well-known criteria~\cite{HZ,ericsteer,Duan,Simon,MDREPR}. In particular, we consider the quadrature based criteria~\cite{Duan,Simon,MDREPR} which were not considered in the earlier paper.

The field of ultra-cold bosons has seen much experimental and theoretical investigation since the successful Bose condensation of bosonic atoms. For atoms trapped in an optical lattice, one investigative technique uses the Bose-Hubbard model. This model, from condensed matter physics, was originally shown by Jaksch~\etal\cite{Jaksch} to provide an accurate description of bosonic atoms trapped in a deep optical lattice. In this work we use a three well Bose-Hubbard model to propose and analyse the entangling properties of a quantum atom optical mode splitter and recombiner. We show that this can split an initial condensate in the central well into two separated entangled condensates, with the detection of the entanglement being sensitive to both the initial quantum state of the condensate in the central well, and to the actual criteria used. We then examine and compare a quantum optical beamsplitter with one vacuum input with regard to the same correlations and input quantum states.

The area of continuous-variable entanglement is very active~\cite{Braunstein,Stefano}, with many criteria having been developed to signify the presence of inseparability and entanglement, especially in bipartite systems. Many of these only apply fully to Gaussian systems and Gaussian measurements. The most commonly used measurements are those developed by Duan \etal\cite{Duan} and Simon~\cite{Simon}, using combinations of quadrature variances. More recently, Teh and Reid have shown the degree of violation of these inequalities that is necessary to demonstrate not just inseparability, but genuine entanglement~\cite{Teh}, as these are only necessarily the same property for pure states.
The criteria we use in this work fall into two categories. Those in the first category were developed by Hillery and Zubairy~\cite{HZ} and expanded on by Cavalcanti \etal\cite{ericsteer} to cover multipartite entanglement, steering, and violations of Bell inequalities. As shown by He \etal\cite{He}, the Hillery and Zubairy criteria are well suited to number conserving processes such as those of interest here. The second category are quadrature based criteria, originally developed by Duan \etal~\cite{Duan} and Simon~\cite{Simon} for inseparability and entanglement, and by Reid~\cite{MDREPR} for demonstrations of the Einstein-Podolsky-Rosen (EPR) paradox~\cite{Einstein}.

Multi-mode entanglement in Bose-Einstein condensates (BEC) has been predicted and examined in the processes of molecular dissociation~\cite{KVK}, four-wave mixing in an optical lattice~\cite{Campbell,Mavis,Andy4}, and in the Bose-Hubbard model~\cite{Hines}. In the latter case the separation of the modes is produced by the tunneling between wells, in both the 
continuous~\cite{Oberthaler2008,Oberthaler2011,He} and pulsed tunneling configurations~\cite{myJPB,myJOSAB}. The quantum correlations necessary to detect entanglement can in principle be measured using the interaction with light~\cite{HomoSimon}, or by homodyning with other atomic modes~\cite{HomoAndy}. 
We note here that the entanglement we are examining is a collective property between atomic modes which are spatially separated, and is not between individual atoms~\cite{Mavis}. This point, which is unavoidable for indistinguishable bosons, has previously been raised by Chianca and Olsen~\cite{SU2Cinthya}, and was recently put on a formal basis, using the language of quantum information theory, by Killoran \etal\cite{Killoran}.

\section{Physical model, Hamiltonian and equations of motion}
\label{sec:model}

We will follow the approach taken by Milburn \etal\cite{BHJoel}, generalising this to three wells~\cite{Nemoto,Chiancathermal}, and solving either the Heisenberg equations of motion or the fully quantum positive-P phase space representation~\cite{Pplus} equations, depending on whether there is a collisional interaction present or not. We consider these to be the most suitable approaches here because they are both exact, allow for an easy representation of mesoscopic numbers of atoms, can be used to calculate quantum correlations, and can simulate different quantum initial states~\cite{states}. Just as importantly, both calculations scale linearly with the number of sites and can in principle deal with any number of atoms. One disadvantage of the positive-P representation is that the integration can show a tendency to diverge at short times for high collisional nonlinearities~\cite{Steel}. As long as the procedures followed to derive the Fokker-Planck equation for the positive-P function are valid~\cite{SMCrispin}, the stochastic solutions are guaranteed to be accurate wherever the integration converges. With all the results shown here, the solutions were found without any signs of divergences.

The system is very simple, with three potential wells in a linear configuration. Each of these can contain a single atomic mode, which we will treat as being in the lowest energy level. Atoms in each of the wells can tunnel into the nearest neighbour potential, with tunneling between wells $1$ and $2$, and $2$ and $3$. With all the population initially in the middle well, the system acts as a time dependent mode splitter and recombiner.  
With the $\hat{a}_{j}$ as bosonic annihilation operators for atoms in mode $j$, $J$ representing the coupling between the wells, and $\chi$ as the collisional nonlinearity, we may now write our Hamiltonian.
Following the usual procedures~\cite{BHJoel}, we find
\begin{equation}
{\cal H} =  \hbar\sum_{j=1}^{3}\chi \hat{a}_{j}^{\dag\;2}\hat{a}_{j}^{2}
+\hbar J\left(\hat{a}_{1}^{\dag}\hat{a}_{2}+\hat{a}_{2}^{\dag}\hat{a}_{1}+\hat{a}_{3}^{\dag}\hat{a}_{2}
+\hat{a}_{2}^{\dag}\hat{a}_{3}\right).
\label{eq:Ham}
\end{equation}

\subsection{Non-interacting case}
\label{subsec:ernest}

For the case where the collisional interaction between the atoms is set to zero, we find that an analytical solution of the Heisenberg equations of motion for the system operators is possible. The Heisenberg equations of motion are found as 
\begin{equation}
\frac{d}{dt}\left[ \begin{array}{c}
\hat{a}_{1} \\ \hat{a}_{1}^{\dag} \\ \hat{a}_{2} \\ \hat{a}_{2}^{\dag} \\ \hat{a}_{3} \\ \hat{a}_{3}^{\dag} \end{array}\right]
=
\left[\begin{array}{cccccc} 
0 & 0 & -iJ & 0 & 0 & 0 \\ 
0 & 0 & 0 & iJ & 0 & 0 \\
-iJ & 0 & 0 & 0 & -iJ & 0 \\
0 & iJ & 0 & 0 & 0 & iJ \\
0 & 0 & -iJ  & 0 & 0 & 0 \\
0 & 0 & 0 & iJ & 0 & 0
\end{array}\right]
\times \left[ \begin{array}{c}
\hat{a}_{1}(0) \\ \hat{a}_{1}^{\dag}(0) \\ \hat{a}_{2}(0) \\ \hat{a}_{2}^{\dag}(0) \\ \hat{a}_{3}(0) \\ \hat{a}_{3}^{\dag}(0)
\end{array} \right].
\label{eq:Heisenberg}
\end{equation} 

This set of linear operator equations is readily solved, having the solutions
\begin{eqnarray}
\eqalign{
\hat{a}_{1}(t) = \frac{1}{2}\left(\cos \Omega t + 1 \right)\hat{a}_{1}(0) - \frac{i}{\sqrt{2}}\sin\Omega t\: \hat{a}_{2}(0) + \frac{1}{2}\left(\cos\Omega t - 1\right)\hat{a}_{3}(0),  \\
\hat{a}_{1}^{\dag}(t) = \frac{1}{2}\left(\cos \Omega t + 1 \right)\hat{a}_{1}^{\dag}(0) + \frac{i}{\sqrt{2}}\sin\Omega t\: \hat{a}_{2}^{\dag}(0) + \frac{1}{2}\left(\cos\Omega t - 1\right)\hat{a}_{3}^{\dag}(0),  \\
\hat{a}_{2}(t) = \frac{-i}{\sqrt{2}}\sin\Omega t\: \hat{a}_{1}(0) + \cos\Omega t\:\hat{a}_{2}(0) - \frac{i}{\sqrt{2}}\sin\Omega t\:\hat{a}_{3}(0),  \\
\hat{a}_{2}^{\dag}(t) = \frac{i}{\sqrt{2}}\sin\Omega t\: \hat{a}_{1}^{\dag}(0) + \cos\Omega t\:\hat{a}_{2}^{\dag}(0) + \frac{i}{\sqrt{2}}\sin\Omega t\:\hat{a}_{3}^{\dag}(0),  \\
\hat{a}_{3}(t) = \frac{1}{2}\left(\cos \Omega t - 1 \right)\hat{a}_{1}(0) - \frac{i}{\sqrt{2}}\sin\Omega t \:\hat{a}_{2}(0) + \frac{1}{2}\left(\cos\Omega t + 1\right)\hat{a}_{3}(0),  \\
\hat{a}_{3}^{\dag}(t) = \frac{1}{2}\left(\cos \Omega t - 1 \right)\hat{a}_{1}^{\dag}(0) + \frac{i}{\sqrt{2}}\sin\Omega t\: \hat{a}_{2}^{\dag}(0) + \frac{1}{2}\left(\cos\Omega t + 1\right)\hat{a}_{3}^{\dag}(0),}
\label{eq:Hsols}
\end{eqnarray}
where we have made the substitution $\Omega = \sqrt{2} J$ for reasons of notational elegance. These equations allow us to find analytical expressions for all the correlations of interest, as we shall do further on in the article. 

We can also solve the Heisenberg equations in terms of $\hat{X}_{i}$ and $\hat{Y}_{i}$, the quadrature operators. Setting
\begin{equation}
\hat{X}_{i} = \hat{a}_{i}+\hat{a}_{i}^{\dag}  \hspace{.5cm} \mbox{and} \hspace{.5cm}
\hat{Y}_{i} = -i\left(\hat{a}_{i}-\hat{a}_{i}^{\dag}\right),
\label{eq:XYdef}
\end{equation}
we find
\begin{eqnarray}
\eqalign{
\hat{X}_{1}(t) = \frac{1}{2}\left(\cos\Omega t + 1\right)\hat{X}_{1}(0)+\frac{1}{\sqrt{2}}\sin\Omega t \: \hat{Y}_{2}(0)+\frac{1}{2}\left(\cos\Omega t -1
\right)\hat{X}_{3}(0), \\
\hat{Y}_{1}(t) = \frac{1}{2}\left(\cos\Omega t + 1\right)\hat{Y}_{1}(0)-\frac{1}{\sqrt{2}}\sin\Omega t \: \hat{X}_{2}(0)+\frac{1}{2}\left(\cos\Omega t -1
\right)\hat{Y}_{3}(0), \\
\hat{X}_{2}(t) = \frac{1}{\sqrt{2}}\sin\Omega t\:\hat{Y}_{1}(0) + \cos\Omega t\:\hat{X}_{2}(0) + \frac{1}{\sqrt{2}}\sin\Omega t\:\hat{Y}_{3}(0), \\
\hat{Y}_{2}(t) = \frac{-1}{\sqrt{2}}\sin\Omega t\:\hat{X}_{1}(0) + \cos\Omega t\:\hat{Y}_{2}(0) - \frac{1}{\sqrt{2}}\sin\Omega t\:\hat{X}_{3}(0), \\
\hat{X}_{3}(t) = \frac{1}{2}\left(\cos\Omega t - 1\right)\hat{X}_{1}(0)+\frac{1}{\sqrt{2}}\sin\Omega t \: \hat{Y}_{2}(0)+\frac{1}{2}\left(\cos\Omega t +1\right)\hat{X}_{3}(0), \\
\hat{Y}_{3}(t) = \frac{1}{2}\left(\cos\Omega t - 1\right)\hat{Y}_{1}(0)-\frac{1}{\sqrt{2}}\sin\Omega t \: \hat{X}_{2}(0)+\frac{1}{2}\left(\cos\Omega t +1\right)\hat{Y}_{3}(0),}
\label{eq:XYernest}
\end{eqnarray}
which then allow us to find solutions for any correlations written in terms of these quadratures. Examples of these are the quadrature squeezing~\cite{DFW} and Reid EPR correlations~\cite{MDREPR}.

\subsection{Interacting case}
\label{subsec:stochastic}

In this case ($\chi\neq 0$), it is not obvious how to solve the equations of motion analytically.
We will therefore use the positive-P representation~\cite{Pplus}, which allows for exact solutions of the dynamics arising from the Hamiltonian of Eq.~\ref{eq:Ham}, in the limit of the average of an infinite number of trajectories of the stochastic differential equations in a doubled phase-space. In practice we obviously cannot integrate an infinite number of trajectories, but have used numbers large enough that the sampling error is within the line thicknesses of our plotted results.
Following the standard methods~\cite{DFW}, the set of It\^o stochastic differential equations~\cite{SMCrispin} are found as
\begin{eqnarray}
\eqalign{
\frac{d\alpha_{1}}{dt} = -2i\chi\alpha_{1}^{+}\alpha_{1}^{2}-iJ\alpha_{2}
+\sqrt{-2i\chi\alpha_{1}^{2}}\;\eta_{1}, \\
\frac{d\alpha_{1}^{+}}{dt} = 2i\chi\alpha_{1}^{+\,2}\alpha_{1}+iJ\alpha_{2}^{+}
+\sqrt{2i\chi\alpha_{1}^{+\;2}}\;\eta_{2}, \\
\frac{d\alpha_{2}}{dt} = -2i\chi\alpha_{2}^{+}\alpha_{2}^{2}-iJ\left(\alpha_{1}
+\alpha_{3}\right)
+\sqrt{-2i\chi\alpha_{2}^{2}}\;\eta_{3}, \\
\frac{d\alpha_{2}^{+}}{dt} = 2i\chi\alpha_{2}^{+\,2}\alpha_{2} + iJ\left(\alpha_{1}^{+}
+\alpha_{3}^{+}\right)
+\sqrt{2i\chi\alpha_{2}^{+\,2}}\;\eta_{4}, \\
\frac{d\alpha_{3}}{dt} = -2i\chi\alpha_{3}^{+}\alpha_{3}^{2}-iJ\alpha_{2}
+\sqrt{-2i\chi\alpha_{3}^{2}}\;\eta_{5}, \\
\frac{d\alpha_{3}^{+}}{dt} = 2i\chi\alpha_{3}^{+\,2}\alpha_{3} + iJ\alpha_{2}^{+}
+\sqrt{2i\chi\alpha_{3}^{+\;2}}\;\eta_{6},}
\label{eq:Pplus}
\end{eqnarray}
where the $\eta_{j}$ are standard Gaussian noises with $\overline{\eta_{j}}=0$ and $\overline{\eta_{j}(t)\eta_{k}(t')}=\delta_{jk}\delta(t-t')$. As always, averages of the positive-P variables represent normally ordered operator moments, such that, for example, $\overline{\alpha_{j}^{m}\alpha_{k}^{+\,n}}\rightarrow\langle\hat{a}^{\dag\,n}\hat{a}^{m}\rangle$. We also note that $\alpha_{j}=(\alpha_{j}^{+})^{\ast}$ only after taking averages, and it is this freedom that allows classical variables to represent quantum operators.

\section{Quantum correlations}
\label{sec:correlations}

\subsection{Analytic solutions}
\label{subsec:analytic}

\begin{figure}
\begin{center}
\includegraphics[width=0.8\columnwidth]{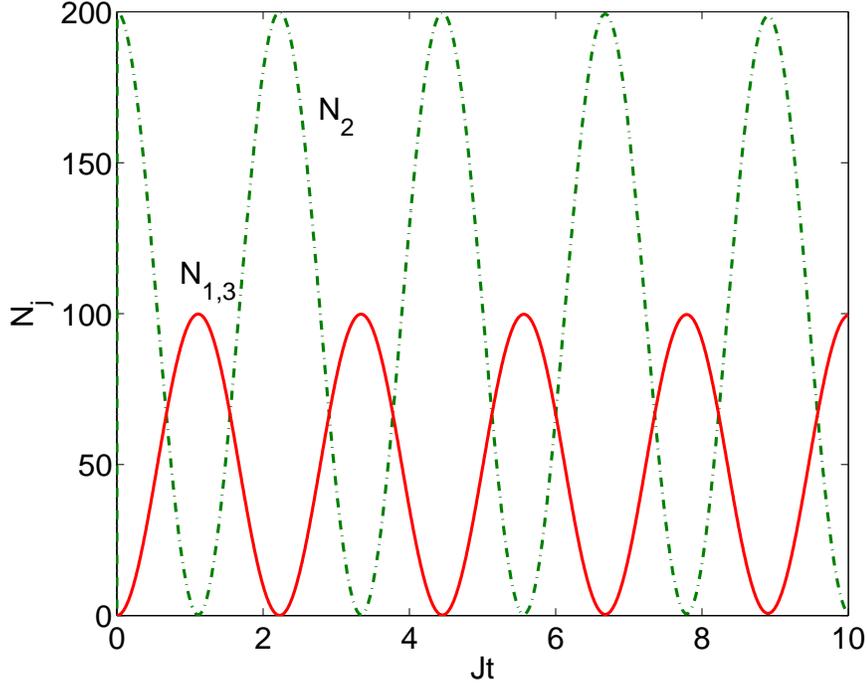}
\end{center}
\caption{(Colour online) The populations in each well as a function of time, for $J=1$, $\chi=10^{-3}$, and $N_{2}(0)=200$, with $N_{1}(0)=N_{3}(0)=0$. The atoms in the centre well begin in a Fock state, although an initial coherent state leads to indistinguishable results. The results shown are the average of $1.08\times 10^{6}$ stochastic trajectories. The non-interacting analytical results are indistinguishable on this scale. The quantities plotted in this and subsequent plots are dimensionless.}
\label{fig:populations}
\end{figure}

As well as the populations in each well, we can also calculate any type of operator products that we desire, analytically in the case without interactions. Beginning with only the middle well occupied, we find the analytic non-interacting solutions for the numbers in each well,
\begin{eqnarray}
\eqalign{
\langle\hat{a}_{1}^{\dag}(t)\hat{a}_{1}(t)\rangle = \langle\hat{a}_{3}^{\dag}(t)\hat{a}_{3}(t)\rangle = \frac{1}{2}\sin^{2}\Omega t \langle\hat{a}_{2}^{\dag}(0)
\hat{a}_{2}(0)\rangle,  \\
\langle\hat{a}_{2}^{\dag}(t)\hat{a}_{2}(t)\rangle = \cos^{2}\Omega t \langle\hat{a}_{2}^{\dag}(0)\hat{a}_{2}(0)\rangle.}
\label{eq:numbers}
\end{eqnarray}
On the scale of Fig.~\ref{fig:populations}, which shows numerical solutions, the above are indistinguishable from the stochastic solutions for $\chi\neq 0$.

\begin{figure}
\begin{center}
\includegraphics[width=0.8\columnwidth,height=0.6\columnwidth]{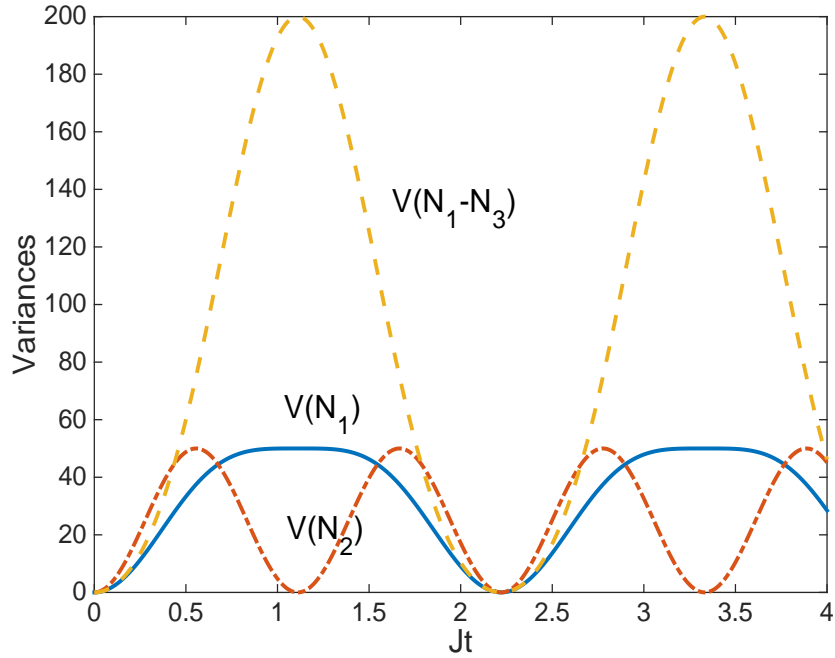}
\end{center}
\caption{(Colour online) The number variances of Eq.~\ref{eq:Heisvars}, for an initial Fock state of $200$ atoms in the middle well. We see that all variances are periodic in the non-interacting case.}
\label{fig:Erwinvar1}
\end{figure}

\begin{figure}
\begin{center}
\includegraphics[width=0.8\columnwidth,height=0.6\columnwidth]{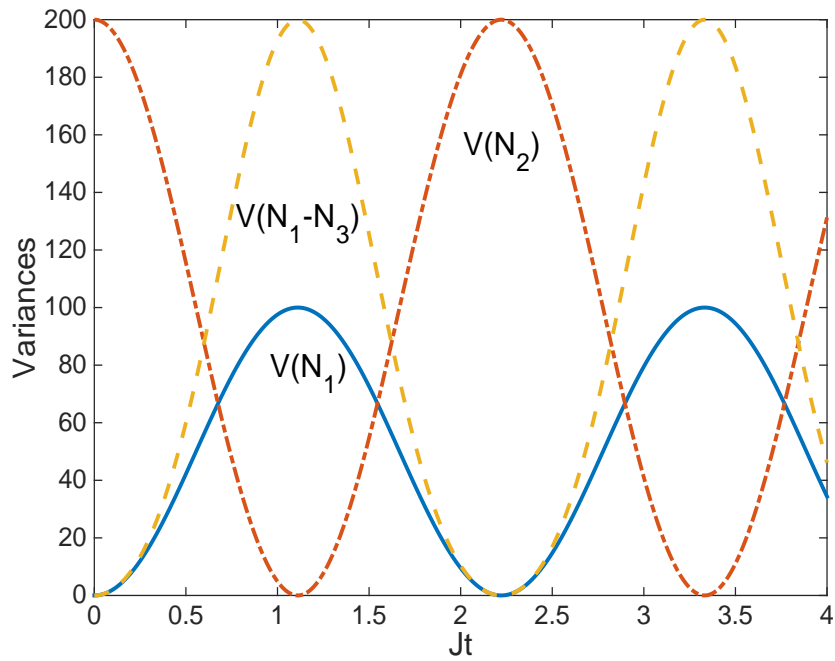}
\end{center}
\caption{(Colour online) The number variances of Eq.~\ref{eq:Heisvars}, for an initial coherent state of $200$ atoms in the middle well. We see that the maximum variances are much larger than in Fig.~\ref{fig:Erwinvar1}, and that all variances are periodic in the non-interacting case.}
\label{fig:Erwinvar2}
\end{figure}

The next class of correlations we calculate are the number variances, including the number difference between the populations of wells $1$ and $3$. In terms of the operators, these are 
\begin{eqnarray}
\eqalign{
V(\hat{N}_{j}) = \langle \hat{a}_{j}^{\dag}\hat{a}_{j}\hat{a}_{j}^{\dag}\hat{a}_{j}\rangle - \langle \hat{a}_{j}^{\dag}\hat{a}_{j}\rangle^{2}, \\
V(\hat{N}_{1}-\hat{N}_{3}) = \langle \left(\hat{a}_{1}^{\dag}\hat{a}_{1}-\hat{a}_{3}^{\dag}\hat{a}_{3}\right)^{2}\rangle - \langle \hat{a}_{1}^{\dag}\hat{a}_{1}-\hat{a}_{3}^{\dag}\hat{a}_{3}\rangle^{2}.}
\label{eq:VNernest}
\end{eqnarray}
In the non-interacting case and with only the middle well initially occupied, we find
\begin{eqnarray}
V(\hat{N}_{1}) &=& V(\hat{N}_{3}) = \frac{1}{4}\left\{\sin^{4}(\Omega t)\;V(\hat{N}_{2}(0))+(1-\cos^{4}(\Omega t))\langle\hat{N}_{2}(0)\rangle\right\}, \nonumber \\
V(\hat{N}_{2}) &=& \cos(\Omega t)^{4}\;V(\hat{N}_{2}(0))+ \frac{1}{4}\sin^{2}2\Omega t \langle \hat{N}_{2}(0)\rangle, \nonumber \\
V(\hat{N}_{1}-\hat{N}_{3}) &=& \frac{1}{2}\sin^{2}\Omega t \left(1+\sin^{2}\Omega t\right)\langle\hat{N}_{2}(0)\rangle.
\label{eq:Heisvars}
\end{eqnarray}
These results, for initial Fock and coherent states in the middle well, are shown in Fig.~\ref{fig:Erwinvar1} and Fig.~\ref{fig:Erwinvar2}.

The second correlation is an entanglement measure adapted from an inequality developed by Hillery and Zubairy, who showed that, considering two separable modes denoted by $i$ and $j$~\cite{HZ},
\begin{equation}
| \langle \hat{a}_{i}^{\dag}\hat{a}_{j}\rangle |^{2} \leq \langle \hat{a}_{i}^{\dag}\hat{a}_{i}\hat{a}_{j}^{\dag}\hat{a}_{j}\rangle,
\label{eq:HZ}
\end{equation}
with the equality holding for coherent states. The violation of this inequality is thus an indication of the inseparability of, and entanglement between, the two modes. Cavalcanti \etal\cite{ericsteer} have extended this inequality to provide indicators of EPR 
steering~\cite{Einstein,Erwin,Wiseman} and Bell violations~\cite{Bell}. We now define the correlation function
\begin{equation}
\xi_{13} = \langle \hat{a}_{1}^{\dag}\hat{a}_{3}\rangle\langle \hat{a}_{1}\hat{a}_{3}^{\dag}\rangle - \langle \hat{a}_{1}^{\dag}\hat{a}_{1}\hat{a}_{3}^{\dag}\hat{a}_{3}\rangle,
\label{eq:xi13}
\end{equation}
for which a positive value reveals entanglement between modes $1$ and $3$. We easily see that $\xi_{13}$ gives a value of zero for two independent coherent states and a negative result for two independent Fock states. This inequality, and the EPR-steering development of it, have been shown to detect both inseparability and asymmetric steering in a three-well Bose-Hubbard model under the process of coherent transfer of atomic population 
(CTAP)~\cite{myJPB,myJOSAB}. In our non-interacting case, with all population initially in well $2$, we find the analytic result
\begin{equation}
\xi_{13} = \frac{1}{4}\sin^{4}\Omega t \left[\langle\hat{N}_{2}(0) \rangle-V(\hat{N}_{2}(0))\right],
\label{eq:HZanalytic}
\end{equation}
so that this measure detects entanglement whenever the initial population in the middle well is in a sub-Poissonian state, with the measure being maximised for a number state. The signature of entanglement identically vanishes for an initial coherent state, which is to be expected since our system is somewhat analogous to a beamsplitter, with linear couplings between the modes~\cite{Killoran}.

Cavalcanti \etal~\cite{ericsteer} further developed the work of Hillery and Zubairy to find inequalities for which the violation denotes the possibility of EPR-steering and Bell states. The EPR-steering inequality for two modes is written as
\begin{equation}
|\langle \hat{a}_{i}\hat{a}_{j}^{\dag}\rangle|^{2} \leq \langle \hat{a}_{i}^{\dag}\hat{a}_{i}(\hat{a}_{j}^{\dag}\hat{a}{j}+\frac{1}{2})\rangle ,
\label{eq:ericEPR}
\end{equation}
while the Bell state inequality is written as
\begin{equation}
|\langle \hat{a}_{i}\hat{a}_{j}^{\dag}\rangle|^{2} \leq \langle (\hat{a}_{i}^{\dag}\hat{a}_{i}+\frac{1}{2})(\hat{a}_{j}^{\dag}\hat{a}{j}+\frac{1}{2})\rangle ,
\label{eq:ericBell}
\end{equation}
Calling on the overworked Alice and Bob, if Alice measures mode $i$ and Bob measures mode $j$ a violation of the inequality (\ref{eq:ericEPR}) signifies that Bob would be able to steer Alice, and vice versa for a swapping of the modes. These inequalities allow us to define a correlation function which signifies the presence of EPR-steering when it has a value of greater than zero,
\begin{equation}
\Sigma_{ij} = \langle \hat{a}_{i}\hat{a}_{j}^{\dag}\rangle\langle \hat{a}_{i}^{\dag}\hat{a}_{j}\rangle - \langle \hat{a}_{i}^{\dag}\hat{a}_{i}(\hat{a}_{j}^{\dag}\hat{a}{j}+\frac{1}{2})\rangle ,
\label{eq:EPRsigma}
\end{equation}
and another for which a positive value signifies the presence of Bell correlations,
\begin{equation}
\zeta_{ij} = \langle \hat{a}_{i}\hat{a}_{j}^{\dag}\rangle\langle \hat{a}_{i}^{\dag}\hat{a}_{j}\rangle - \langle (\hat{a}_{i}^{\dag}\hat{a}_{i}+\frac{1}{2})(\hat{a}_{j}^{\dag}\hat{a}{j}+\frac{1}{2})\rangle .
\label{eq:Bellbeta}
\end{equation}

For the EPR-steering correlation, we can solve the Heisenberg equations to find
\begin{equation}
\Sigma_{13} = \Sigma_{31} = \frac{1}{4}\sin^{2}\Omega t \left(\sin^{2}\Omega t-1\right)\langle \hat{N}_{2}(0)\rangle-\frac{1}{2}V(\hat{N}_{2}(0))\sin^{4}\Omega t,
\label{eq:ernestSigma}
\end{equation}
which is readily seen to have a maximum value of zero. Therefore this measure does not detect any possibility of EPR-steering for this system. There will also obviously be no signature of a continuous variable Bell state of the two modes.

One other common method of detecting entanglement in continuous variable systems involves using quadrature correlations~\cite{Duan,Simon}. The single-mode quadrature variances for the non-interacting case can be found as
\begin{eqnarray}
\eqalign{
\hspace{-2cm} V(\hat{X}_{1}(t)) = \frac{1}{4}\left(\cos\Omega t+1\right)^{2}V(\hat{X}_{1}(0))+\frac{1}{2}\sin^{2}\Omega t\: V(\hat{Y}_{2}(0))+\frac{1}{4}\left(\cos\Omega t-1\right)^{2}V(\hat{X}_{3}(0)), \nonumber \\
\hspace{-2cm} V(\hat{Y}_{1}(t)) = \frac{1}{4}\left(\cos\Omega t+1\right)^{2}V(\hat{Y}_{1}(0))+\frac{1}{2}\sin^{2}\Omega t\: V(\hat{X}_{2}(0))+\frac{1}{4}\left(\cos\Omega t-1\right)^{2}V(\hat{Y}_{3}(0)), \nonumber \\
\hspace{-2cm} V(\hat{X}_{2}(t)) = \frac{1}{2}\sin^{2}\Omega t\: V(\hat{Y}_{1}(0))+\cos^{2}\Omega t \:V(\hat{X}_{2}(0))+\frac{1}{2}\sin^{2}\Omega t \: V(\hat{Y}_{3}(0)), \nonumber \\
\hspace{-2cm} V(\hat{Y}_{2}(t)) = \frac{1}{2}\sin^{2}\Omega t \:V(\hat{X}_{1}(0))+\cos^{2}\Omega t \:V(\hat{Y}_{2}(0))+\frac{1}{2}\sin^{2}\Omega t \: V(\hat{X}_{3}(0)), \nonumber \\
\hspace{-2cm} V(\hat{X}_{3}(t)) = \frac{1}{4}\left(\cos\Omega t-1\right)^{2}V(\hat{X}_{1}(0)+\frac{1}{2}\sin^{2}\Omega t \: V(\hat{Y}_{2}(0))+\frac{1}{4}\left(\cos\Omega t+1\right)^{2}V(\hat{X}_{3}(0)), \nonumber \\
\hspace{-2cm} V(\hat{Y}_{3}(t)) = \frac{1}{4}\left(\cos\Omega t-1\right)^{2}V(\hat{Y}_{1}(0))+\frac{1}{2}\sin^{2}\Omega t \: V(\hat{X}_{2}(0))+\frac{1}{4}\left(\cos\Omega t+1\right)^{2}V(\hat{Y}_{3}(0)).}
\label{eq:VXY}
\end{eqnarray}
In experimental quantum optics, these quadrature variances are measured via homodyne detection, which is not as simple for massive particles, although at least two methods have been proposed~\cite{HomoSimon,HomoAndy}.

We now investigate the Duan-Simon correlations between wells $1$ and $3$, with inseparability being detected when
\begin{equation}
V(\hat{X}_{1}\pm\hat{X}_{3})+V(\hat{Y}_{1}\mp \hat{Y}_{3}) < 4,
\label{eq:DSformula}
\end{equation}
where for simplicity of expression we have dropped the time variable.
To express these particular correlations, we also need the quadrature covariances, which are found as
\begin{eqnarray}
\eqalign{
\hspace{-1cm} V(\hat{X}_{1},\hat{X}_{3}) = \frac{1}{4}\left(\cos^{2}\Omega t-1\right)\left[V(\hat{X}_{1}(0))+V(\hat{X}_{3}(0))\right]+\frac{1}{2}\sin^{2}\Omega t\:V(\hat{Y}_{2}(0)), \\
\hspace{-1cm} V(\hat{Y}_{1},\hat{Y}_{3}) = \frac{1}{4}\left(\cos^{2}\Omega t-1\right)\left[V(\hat{Y}_{1}(0))+V(\hat{Y}_{3}(0))\right]+\frac{1}{2}\sin^{2}\Omega t\:V(\hat{X}_{2}(0)).}
\label{eq:covariances}
\end{eqnarray}
Setting $DS_{\pm}=V(\hat{X}_{1}\pm\hat{X}_{3})+V(\hat{Y}_{1}\mp\hat{Y}_{3})$, we find
\begin{eqnarray}
\eqalign{
\hspace{-2cm} DS_{+} = V(\hat{Y}_{1}(0))+V(\hat{Y}_{3}(0))+\cos^{2}\Omega t\left[V(\hat{X}_{1}(0))+V(\hat{X}_{3}(0))\right]+2\sin^{2}\Omega t\: V(\hat{Y}_{2}(0)), \\
\hspace{-2cm} DS_{-} = V(\hat{X}_{1}(0))+V(\hat{X}_{3}(0))+\cos^{2}\Omega t\left[V(\hat{Y}_{1}(0))+V(\hat{Y}_{3}(0))\right]+2\sin^{2}\Omega t\: V(\hat{X}_{2}(0)), }
\label{eq:DSresults}
\end{eqnarray}
which can then be minimised with respect to time. We find that the minimum for either correlation is at $4$, so that inseparability for our initial conditions is not found by this measure, irrespective of the initial quantum state of the atoms in well $2$. We also found that the Reid EPR inequalities~\cite{MDREPR} showed no evidence of the EPR paradox. This is despite the fact that entanglement is present according to the Hillery-Zubairy criteria~\cite{HZ}, and emphasises the importance of using the correct correlations to detect continuous-variable entanglement in a given system.

\begin{figure}
\begin{center}
\includegraphics[width=0.8\columnwidth]{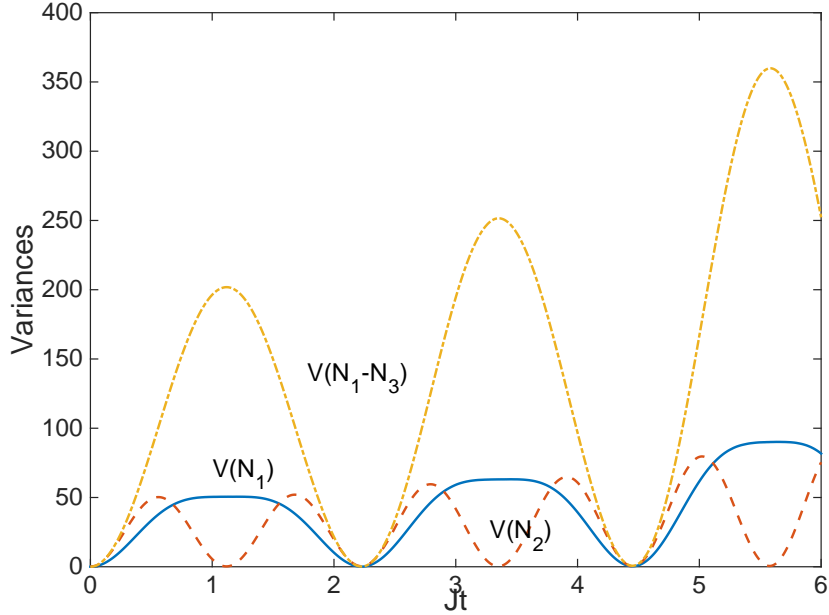}
\end{center}
\caption{(Colour online) The number variances as a function of time, for $J=1$, $\chi=10^{-3}$, and $N_{2}(0)=200$ in a Fock state, with $N_{1}(0)=N_{3}(0)=0$. The results shown are the average of $1.69\times 10^{6}$ stochastic trajectories. The non-interacting analytical results are only the same at short times and we see that the amplitudes of the oscillations grow with time, despite the fact that a $\chi^{(3)}$ nonlinearity preserves the number statistics in an isolated well.}
\label{fig:PPvars}
\end{figure}

\subsection{Numerical solutions}
\label{subsec:numerics}

In the interacting case, our method of choice is to use numerical stochastic integration to find solutions of the full positive-P representation equations~\cite{CVCandme}. This allows us to calculate the expectation values of any operator moments that can be written in normal order.  
Taking into account the normal ordering, the number variances are written as
\begin{eqnarray}
\eqalign{
V(\hat{N}_{j}) = \overline{ \alpha_{j}^{+\,2}\alpha_{j}^{2}}+\overline{\alpha_{j}^{+}\alpha_{j}}-\overline{\alpha_{j}^{+}\alpha_{j}}^{2}, \\
V(\hat{N}_{1}-\hat{N}_{3}) = V(\hat{N}_{1})+V(\hat{N}_{3})-2V(\hat{N}_{1},\hat{N}_{3}), \\
= V(\hat{N}_{1})+V(\hat{N}_{3})-2\left(\overline{\alpha_{1}^{\dag}\alpha_{1}\alpha_{3}^{\dag}\alpha_{3}} -\overline{\alpha_{1}^{+}\alpha_{1}}\times\overline{\alpha_{3}^{+}\alpha_{3}}\right).}
\label{eq:variances} 
\end{eqnarray}
All of these give values of zero for uncorrelated Fock states or vacuum. Whenever one of the variances is less than the mean population of the corresponding mode, we have suppression of number fluctuations below the Poissonian coherent state level. 
The individual quadrature variances are found as
\begin{eqnarray}
\eqalign{
V(\hat{X}_{i}) = 1+2\overline{\alpha_{i}^{+}+\alpha_{i}^{2}+\alpha_{i}^{+\;2}}-\overline{\alpha_{i}+\alpha_{i}^{+}}^{2}, \\
V(\hat{Y}_{i}) = 1+2\overline{\alpha_{i}^{+}-\alpha_{i}^{2}-\alpha_{i}^{+\;2}}-\overline{-i(\alpha_{i}-\alpha_{i}^{+})}^{2},}
\label{eq:Pplusquads}
\end{eqnarray}
with the combined quadrature variances and covariances needed for the Duan-Simon and Reid correlations being the obvious extensions of these.

\begin{figure}
\begin{center}
\includegraphics[width=0.8\columnwidth]{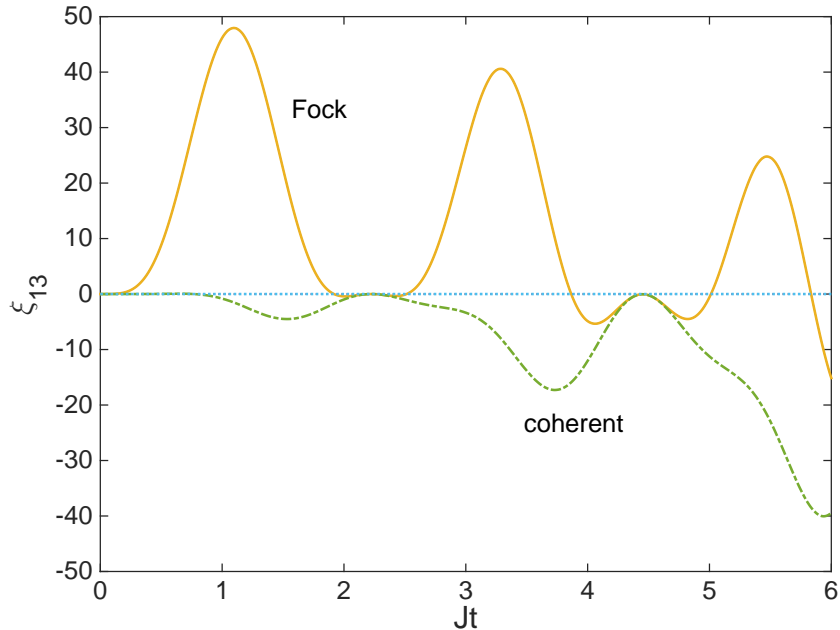}
\end{center}
\caption{(Colour online) The Hillery-Zubairy criteria as a function of time, for $J=1$, $\chi=10^{-3}$, and $N_{2}(0)=200$ in initial Fock state and coherent states, with $N_{1}(0)=N_{3}(0)=0$. The results shown for the initial Fock state are the average of $1.69\times 10^{6}$ stochastic trajectories and those for the initial coherent state are averaged over $1.25\times 10^{6}$ trajectories. The non-interacting analytical results are only the same at short times and we see that the correlations degrade with time, with the coherent state correlation showing no entanglement at any time.}
\label{fig:HZPP}
\end{figure}

For our results in the interacting case, we have chosen a nonlinearity of $\chi=10^{-3}$, again with either a Fock or coherent state with an average of $200$ atoms in the middle well. These different quantum states are simulated using the methods found in Olsen and Bradley~\cite{states}. We have simulated results for the numbers in each well (Fig.~\ref{fig:populations}), the number variances (Fig.~\ref{fig:PPvars}), the Hillery-Zubairy criteria (Fig.~\ref{fig:HZPP}) and some of the various quadrature variance correlations canonically used to detect continuous-variable entanglement and EPR-steering. Those we present here are the Duan-Simon criteria~\cite{Duan,Simon} of Eq.~\ref{eq:DSformula} and the Reid EPR inequalities~\cite{MDREPR}. 

 \begin{figure}
\begin{center}
\includegraphics[width=0.8\columnwidth]{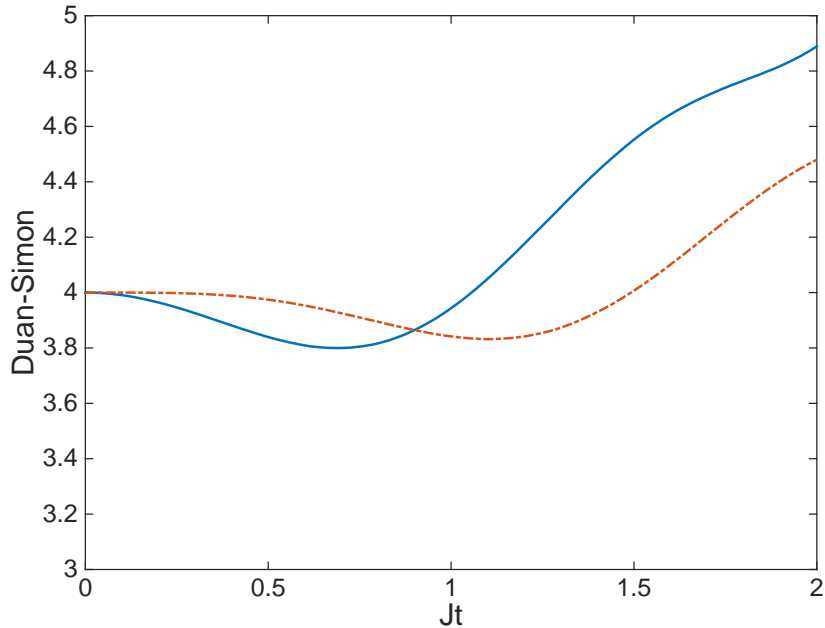}
\end{center}
\caption{(Colour online) The Duan-Simon inseparability criteria as a function of time, for $J=1$, $\chi=10^{-3}$, and $N_{2}(0)=200$ in a coherent states, with $N_{1}(0)=N_{3}(0)=0$. The solid line is $V(\hat{X}_{1}-\hat{X}_{2})+V(\hat{Y}_{1}+\hat{Y}_{2})$ and the dash-dotted line is $V(\hat{X}_{1}-\hat{X}_{3})+V(\hat{Y}_{1}+\hat{Y}_{3})$ We see that inseparability is only indicated for short times and that the violation of the inequality is not large. These results are the average of $9.45\times 10^{5}$ trajectories.}
\label{fig:DScoherent}
\end{figure}

The number variances for an initial Fock state are shown in Fig.~\ref{fig:PPvars}, from which we can see that they take the same periodic form as in the non-interacting case of Fig.~\ref{fig:Erwinvar1}, but the amplitude of the oscillations grows in time. The results for an initial coherent state follow the same pattern as in Fig.~\ref{fig:Erwinvar2}, but again with the maxima of the oscillations increasing with time.  The increase in these variances is purely a result of the linear coupling between the wells, since the collisional nonlinearity in an isolated mode preserves the number statistics. Although of the same strengths, the coupling between the wells is independent, so that we see the statistics of $\hat{N}_{1}-\hat{N}_{3}$ are initially Poissonian. The interaction of the collisional nonlinearity and the couplings causes the statistics to become super-Poissonian with increasing interaction time.

In Fig.~\ref{fig:HZPP} we show the Hillery-Zubairy criterion $\xi_{13}$ for the detection of entanglement between wells $1$ and $3$. We see that an initial Fock state in the centre well means that this correlation becomes periodically positive at early times, but that the entanglement signal is degraded over time. For an initial coherent state, this measure gives no indication of entanglement. This is in contradiction with the results of Fig.~\ref{fig:DScoherent} where consider an initial coherent state and find a violation of Duan-Simon inequalities at short times. We note that using time dependent quadrature angles can maximise the violations, as shown previously for Kerr-squeezed optical states mixed on a beamsplitter~\cite{nlc,JoelKerr}, but we have not considered this here since the inseparability signal for the canonical quadratures is so weak. In any case, even an optimisation of the quadrature angles still finds no violation of the inequalities after a short time. We also calculated the Reid EPR criteria between wells $1$ and $2$ and $1$ and $3$, and found no evidence that EPR-steering is present in this system. In this case, the quadrature measures agree with the phase-independent measures of Eq.~\ref{eq:ericEPR}.

\section{The beamsplitter}
\label{sec:BS}

Since our three well system, with one input mode and two output modes with linear couplings, can loosely be compared to a beamsplitter with one non-zero input, it is informative to compare the performance of a standard optical beamsplitter using the same correlations. The equations relating the inputs and outputs of a lossless beamsplitter can be written as
\begin{eqnarray}
\eqalign{
\hat{a}_{out} = \sqrt{\eta}\:\hat{a}_{in}+ \sqrt{1-\eta}\:\hat{b}_{in}, \\
\hat{b}_{out} = -\sqrt{1-\eta}\:\hat{a}_{in} + \sqrt{\eta}\:\hat{b}_{in},}
\label{ernieBS}
\end{eqnarray}
where $\sqrt{\eta}$ is the amplitude reflectivity. For reasons of simplicity, we will treat only a balanced beamsplitter, with $\eta=1/2$ and $\langle\hat{b}_{in}^{\dag}\hat{b}_{in}\rangle = 0$.  It is then trivial to see that total number is conserved, as it must be, and as happens for our three well system. One obvious difference is that the beamsplitter has two inputs and two outputs and the transmission is not time dependent, but our main interest here is in the linear coupling. This is present in both systems.

Examining firstly the correlation of Eq.\ref{eq:xi13}, we find that the left hand side is
\begin{equation}
\langle \hat{a}_{out}^{\dag}\hat{b}_{out}\rangle\langle\hat{b}^{\dag}_{out}\hat{a}_{out}\rangle = \frac{1}{4}\langle\hat{a}_{in}^{\dag}\hat{a}_{in}\rangle^{2},
\label{eq:adbbdaBS}
\end{equation}
and the right hand side of the expression is
\begin{equation}
\langle \hat{a}^{\dag}_{out}\hat{a}_{out}\hat{b}^{\dag}_{out}\hat{b}_{out}\rangle = \frac{1}{4}\left[\langle(\hat{a}_{in}^{\dag}\hat{a}_{in})^{2}\rangle - \langle\hat{a}_{in}^{\dag}\hat{a}_{in}\rangle
\right].
\label{eq:NaNbBS}
\end{equation}
Combining these, we find
\begin{equation}
\xi_{ab} = \frac{1}{4}\left[\langle\hat{a}_{in}^{\dag}\hat{a}_{in}\rangle-V(\hat{N}\!a_{in})
\right],
\label{eq:xiabBS}
\end{equation} 
which, apart from the time dependence, is the same as the result of Eq.~\ref{eq:HZanalytic}. This again shows that this measure will detect entanglement for any input state $a$ for which the number fluctuations are less than Poissonian. Two coherent states will give a value of zero, and therefore will not lead to entangled outputs. The two possible equivalents of Eq.~\ref{eq:ericEPR} for the beamsplitter give
\begin{equation}
\Sigma_{ab} = \Sigma_{ba} = -V(\hat{N}\!a_{in}),
\label{eq:abSigma}
\end{equation}
which can obviously never be positive, so that these measures do not detect EPR-steering.

We can also consider the Duan-Simon quadrature correlations~\cite{Duan,Simon} using the same approach. We write the output quadratures in terms of the inputs as
\begin{eqnarray}
\eqalign{
\hat{X}_{a}^{out} = \sqrt{\eta}\:\hat{X}_{a}^{in}+\sqrt{1-\eta}\:\hat{X}_{b}^{in},  \\
\hat{Y}_{a}^{out} = \sqrt{\eta}\:\hat{Y}_{a}^{in}+\sqrt{1-\eta}\:\hat{Y}_{b}^{in}, \\
\hat{X}_{b}^{out} = \sqrt{\eta}\hat{X}_{b}^{in}-\sqrt{1-\eta}\:\hat{X}_{a}^{in},  \\
\hat{Y}_{b}^{out} = \sqrt{\eta}\:\hat{Y}_{b}^{in}-\sqrt{1-\eta}\:\hat{Y}_{a}^{in},}
\label{eq:XYernieBS}
\end{eqnarray}        
which alllows us to calculate the necessary quadrature moments analytically. 

Again for simplicity, we set $\eta=1/2$, and find 
\begin{eqnarray}
\eqalign{
V(\hat{X}_{a}^{out}\pm \hat{X}_{b}^{out}) = V(\hat{X}_{a}^{in})+V(\hat{X}_{b}^{in})\pm\left[V(\hat{X}_{b}^{in})-V(\hat{X}_{a}^{in})
\right],  \\
V(\hat{Y}_{a}^{out}\mp \hat{Y}_{b}^{out}) = V(\hat{Y}_{a}^{in})+V(\hat{Y}_{b}^{in})\mp\left[V(\hat{Y}_{b}^{in})-V(\hat{Y}_{a}^{in})
\right],}
\label{eq:VXYpm}
\end{eqnarray}
so that the Duan-Simon correlations are
\begin{eqnarray}
\eqalign{
V(\hat{X}_{a}^{out}+ \hat{X}_{b}^{out}) + V(\hat{Y}_{a}^{out}- \hat{Y}_{b}^{out}) = 2\left[V(\hat{X}_{b}^{in})+V(\hat{Y}_{a}^{in})\right], \\
V(\hat{X}_{a}^{out}- \hat{X}_{b}^{out}) + V(\hat{Y}_{a}^{out}+ \hat{Y}_{b}^{out}) = 2\left[V(\hat{X}_{a}^{in})+V(\hat{Y}_{b}^{in})\right].}
\label{eq:BSDuanSimon}
\end{eqnarray}
For a squeezed amplitude input in mode $a$ with variance $V(\hat{X}_{a}^{in})=\mbox{e}^{-r}$ and vacuum in $b$, the second of these gives a value of $2(1+\mbox{e}^{-r})$, therefore demonstrating inseparability.
However, for an input Fock state in mode $a$ and vacuum in $b$, these correlations predict a value of $4N\!a_{in}+4$, immediately contradicting the prediction of Eq.~\ref{eq:xiabBS}, showing once again the importance of using the correct inequalities for a given system.

Using these analytic results, we can also find values for the Reid EPR correlations~\cite{MDREPR}, for which
\begin{equation}
V^{inf}(\hat{X}_{j}^{out})V^{inf}(\hat{Y}_{j}^{out}) < 1
\label{eq:MDREPRBS}
\end{equation}
signifies a demonstration of the EPR paradox~\cite{Einstein}. Calling the product of the inferred variances $\Gamma_{j}$, we find for inputs of a squeezed state and vacuum,
\begin{equation}
\Gamma_{a}=\Gamma_{b} = \frac{2}{1+\cosh r},
\label{eq:EPRBSsqueeze}
\end{equation}
showing that the paradox is demonstrated and steering is possible as soon as we have a squeezed input. On the other hand, for inputs of a Fock state $|N\rangle$ and vacuum, we find
\begin{equation}
\Gamma_{a}=\Gamma_{b} = \frac{(N^{2}+4N+2)^{2}}{4N^2+4},
\label{eq:EPRBSFock}
\end{equation}
which has a minimum value of $1$. We therefore see that the Reid measure does not signify the presence of the EPR paradox in this case, in agreement with Eq.~\ref{eq:abSigma}.

\section{Conclusions}
\label{sec:conclusions}

We have shown that our three-well Bose Hubbard system produces entanglement between the atoms in two non-adjacent wells, but not at a sufficient level to demonstrate the EPR paradox via the measures we have investigated here. As we have demonstrated, different measures can lead to different indications as to whether inseparability and entanglement are present. Those based on the Hillery-Zubairy results perform better for an initial Fock state than for an initial coherent state in the middle well, for which the quadrature based correlations have a superior performance. This is in agreement with the claim that the Hillery-Zubairy measures are superior for processes which conserve number. Due to the sufficient but not necessary nature of the inequalities used, we cannot say that a demonstration of EPR-steering is impossible with this system, only that we have not found evidence for one. This is an ongoing problem with continuous-variable quantum information, with no single method adequately capturing all the quantum correlations that may exist in a given system.

We have also compared the performance of our system to an optical beamsplitter with one vacuum input, finding some similarities and some differences. The Hillery-Zubairy criteria again predict entanglement for an input Fock state into the bright input, but no EPR-steering. The quadrature correlations predict neither entanglement nor EPR-steering for an initial Fock state, but predict both for squeezed states. The biggest difference is that the beamsplitter outputs do not depend on time, at least for continuous inputs, while the atomic system experiences either periodic or almost periodic behaviour, depending on the presence or otherwise of atomic collisions. As long as the interactions are not too strong, our system is a good proposal for the manufacture of entangled bosonic modes of separated atoms.

\ack

This research was supported by the Australian Research Council under the Future Fellowships Program (Grant ID: FT100100515).

\end{document}